\documentstyle[aps,prl,epsf]{revtex}

\begin{document}

\twocolumn[\hsize\textwidth\columnwidth\hsize\csname
@twocolumnfalse\endcsname
\title{Berry phases in superconducting transitions}
\author{A.A. Aligia}
\address{Centro At\'{o}mico Bariloche and Instituto Balseiro, Comisi\'{o}n Nacional\\
de Energ{\'{\i }}a At\'{o}mica, 8400 S.C. de Bariloche, Argentina.}
\date{Received \today }
\maketitle

\begin{abstract}
I generalize the concept of Berry's geometrical phase for quasicyclic
Hamiltonians to the case in which the ground state evolves adiabatically to
an excited state after one cycle, but returns to the ground state after an
integer number of cycles. This allows to extend the charge Berry phase 
$\gamma _{c}$ related to the macroscopic polarization, to many-body systems
with fractional number of particles per site. Under certain conditions, 
$\gamma _{c}$ and the spin Berry phase $\gamma _{s}$ jump in $\pi $ at the
boundary of superconducting phases. In the extended Hubbard chain with
on-site attraction $U$ and nearest-neighbor interaction $V$ at quarter
filling, the transitions detected agree very well with exact results in two
limits solved by the Bethe ansatz, and with previous numerical studies. In
chains with spin SU(2) symmetry, $\gamma _{s}$ jumps when a spin gap opens.



\end{abstract}

\vskip2pc]

\narrowtext

In the last decade, Berry phases have caused a great deal of interest in a
variety of fields in physics. Applications of this concept to condensed
matter began with the study of Zak of the dynamics of a Bloch electron as
its wave vector {\bf k} changes adiabatically due to an external
perturbation until it reaches {\bf k+G} where {\bf G} is a reciprocal
lattice vector \cite{zak1}. Although in this case the Hamiltonian which
describes the evolution of the periodic part of the Bloch function is
noncyclic, the initial and final points are related by a gauge
transformation and a Berry phase can be defined for each band \cite{zak2},
which describes the center of gravity of the density of the Wannier
function, or Wyckoff position \cite{zak1,mic}. Later progress, showed that
changes in the macroscopic polarization of a band insulator in an
independent particle approximation, are proportional to changes in a Berry
phase, which I call $\gamma _{c}$ \cite{kin,res1}. The formalism was
generalized to the many-body case \cite{or1,or2} extending previous
derivations for quantized charge transport \cite{niu}, and was applied to
study a ferroelectric transition in a strongly correlated model \cite{res2}.

Recently Gagliano and me have introduced the spin Berry phase $\gamma _{s}$,
and have calculated $\gamma _{c}$ and $\gamma _{s}$ for the first time in a
gapless metallic phase \cite{ali1,ali2}. The model was an extended Hubbard
chain with correlated hopping, and due to the presence of inversion symmetry
in it, $\gamma _{c}$ and $\gamma _{s}$ are quantized and can only have the
values 0 and $\pi $. The topological quantum numbers $\gamma _{c}/\pi $ and 
$\gamma _{s}/\pi $ were used as order parameters to construct a phase
diagram, separating the three phases of the model: charge-density wave
(CDW), spin-density wave (SDW) and the metallic phase. The presence of
anomalous flux quantization \cite{ali1}, and additional arguments \cite{ali2}, 
suggest that this phase is superconducting (S), possibly of triplet
character. This work opened the possibility to study phase diagrams in
strongly-correlated systems, by looking at topological transitions, which
are sharp even in systems of finite size. However, all calculations of 
$\gamma _{c}$ and $\gamma _{s}$ in many-body systems, were so far restricted
to an integer number of particles per site. This restriction is too severe,
particularly if one is interested in superconductors.

In this Letter, I extend the definition of the Berry phase to the case in
which it is necessary to perform more than one cycle in a parameter space,
while the Hamiltonian ends at a point related to the starting one by a gauge
transformation, in order for the system to return to a state equivalent to
the initial ground state. At first sight, this seems to contradict the
adiabatic theorem, since there should be a crossing of energy levels before
the first cycle ends. However, this crossing is not essential if the matrix
element of the Hamiltonian between the states involved vanishes, as it is
the case if both states differ in some quantum number. While it seems
difficult to imagine a quantum-mechanical system with these properties, the
generalization below of the charge Berry phase $\gamma _{c}$ to systems with
fractional number of particles per site is an example. Calculating $\gamma
_{c}$ numerically in a system of 12 sites, I obtain the phase diagram of the
extended Hubbard chain for $1/2$ particles per site, negative $U$ and any $V$. 
The results are compared with previous numerical ones, and with analytical
ones obtained mapping the model in the limits $|U|\gg t,|V|$ and $|U|,V\gg t$
to Bethe ansatz solvable cases through appropriate canonical
transformations. In one dimension (1D), for SU(2) symmetric systems, I show
that the spin Berry phase jumps at the boundary between dominating singlet
and triplet correlations at large distances, for any filling.

For simplicity I restrict the discussion to 1D and one band. Extension to
the general case is straightforward. I consider a system of $L$ sites and
number of particles $N$ with $N/L=n/l$, and $n/l$ irreducible. The creation
operators satisfy arbitrary boundary conditions for both spins 
${\bar{c}}_{j+L\sigma }^{\dagger }=
e^{i\Phi _{\sigma }}{\bar{c}}_{j\sigma }^{\dagger }$, 
and the Hamiltonian $\bar{H}(\Phi _{\uparrow },\Phi _{\downarrow })$
conserves number of particles, $z$ component of the total spin, and is
invariant under translations. This invariance allows to define weighted
irreducible representations of the translation group characterized by the
total wave vector $\bar{K}=K+(N_{\uparrow }\Phi _{\uparrow }+N_{\downarrow
}\Phi _{\downarrow })/L$, where $N_{\sigma }$ is the number of particles
with spin $\sigma $ and 
$K=\mathop{\rm integer} \times 2\pi /L$ 
is one of the allowed wave vectors for periodic boundary
conditions (PBC) \cite{afq}. Using the gauge transformation $c_{j\sigma
}^{\dagger }=e^{-ij\Phi _{\sigma }/L}{\bar{c}}_{j\sigma }^{\dagger }$, the
Hamiltonian is converted into one in which the fluxes $\Phi _{\sigma }$ are
distributed equally in each link $H(\Phi _{\uparrow },\Phi _{\downarrow })$
(the hoppings acquire a phase \cite{ali1}), with creation operators
satisfying PBC ${c}_{j+L\sigma }^{\dagger }={c}_{j\sigma }^{\dagger }$. Note
that while $\bar{H}(0,0)=\bar{H}(\pm 2\pi ,\pm 2\pi )$, 
$H(0,0)=\bar{H}(0,0)\neq H(\pm 2\pi ,\pm 2\pi )$ \cite{zak2}.
To generalize $\gamma _{c}$
($\gamma _{s}$), the idea is to start from the ground state $|g_{K}(\Phi
_{\uparrow },\Phi _{\downarrow })\rangle $ of $H(\Phi _{\uparrow },\Phi
_{\downarrow })$ for wave vector $K$ and fluxes $\Phi _{\sigma }$ which
minimize the energy, and follow the phase of $|g_{K}\rangle $ as the $\Phi
_{\sigma }$ are shifted adiabatically by the same amount, with the same
(opposite) sign, until $|g_{K}\rangle $ reaches a state equivalent to the
initial one (the same eigenstate of $\bar{H}$ except for a phase). I discuss
first $\gamma _{c}$. It can be defined as $\gamma _{c}= i \int {}_{0}^{2\pi
l}d\Phi \langle g_{K}(\Phi ,\Phi )|\frac{\partial }{\partial \Phi }
g_{K}(\Phi ,\Phi )\rangle $, or in the numerically gauge invariant form,
discretizing the interval $0\leq \Phi \leq 2\pi l$ into $M$ points $\Phi
_{i}=2\pi li/M$: 

\begin{eqnarray}
{\small \gamma }_{c} &=&{\small -}\lim_{M\rightarrow \infty }\text{Im}\ln 
{\small \Pi }_{i=0}^{M-2}{\small \langle g}_{K}{\small (\Phi }_{i},{\small %
\Phi }_{i}{\small )|g}_{K}{\small (\Phi }_{i+1}{\small ,\Phi }_{i+1}{\small %
)\rangle }  \nonumber \\
&&\times {\small \newline
\langle g}_{K}{\small (\Phi }_{M-1}{\small ,\Phi }_{M-1}{\small )|}e^{%
{\small i}\frac{{2\pi l}}{L}\sum_{j\sigma }{\small jc}_{j\sigma }^{\dagger }%
{\small c}_{j\sigma }}{\small |g}_{K}{\small (0,0)\rangle }  \label{f1}
\end{eqnarray}
\noindent where the last ket is $|g_{K}(2\pi l,2\pi l)\rangle $
calculated using the gauge transformation from the eigenstate 
$|g_{K}(0,0)\rangle $ of $\bar{H}(0,0)$. In practice, $\gamma _{c}$ can be
calculated to four digits accuracy using $M\sim 6$ adequately chosen points.

\begin{figure}
\narrowtext
\epsfxsize=3.0truein
\vbox{\hskip 0.05truein \epsffile{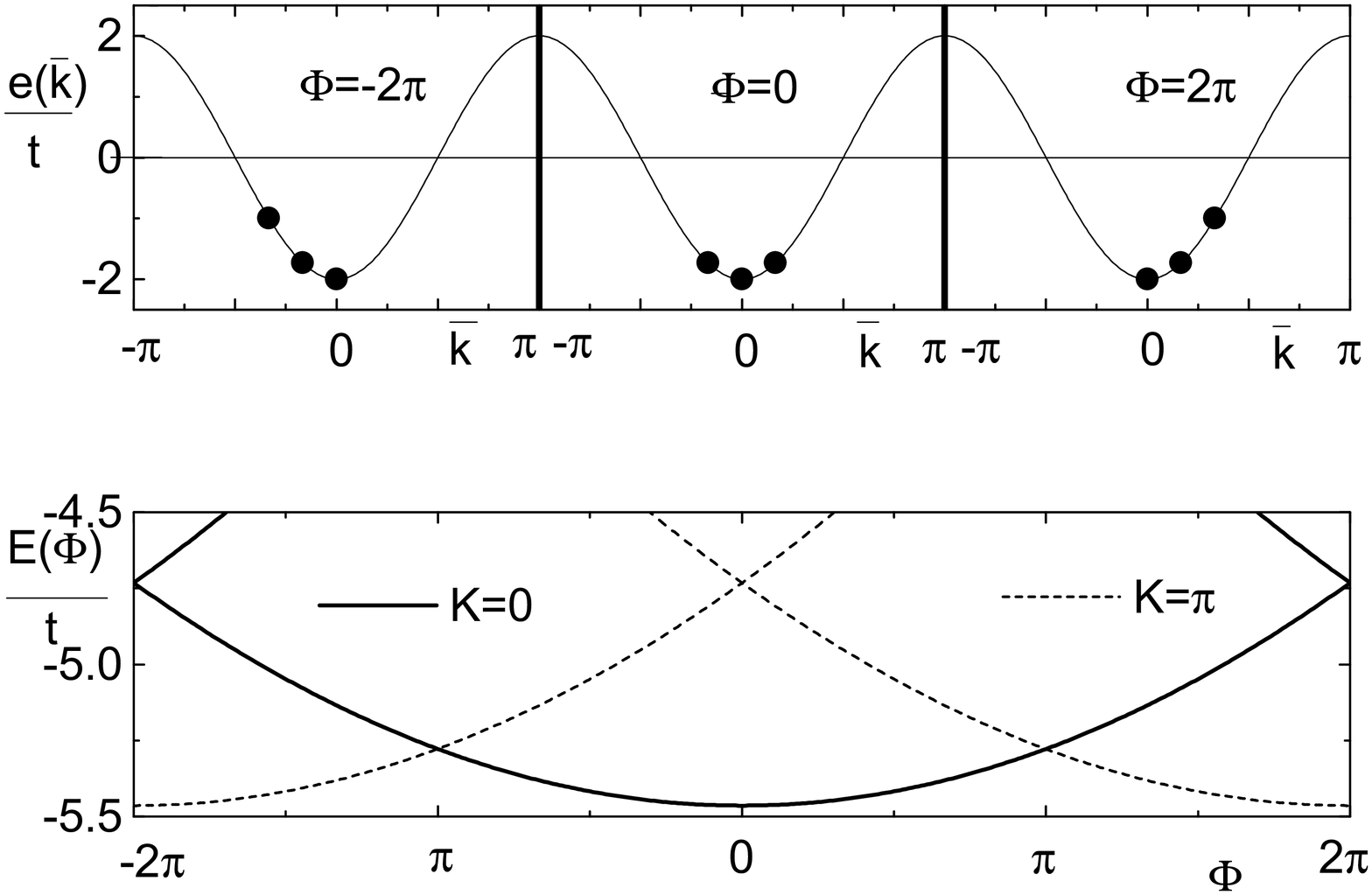}}
\medskip
\caption{Top: Scheme of the evolution of the ground state with flux $\Phi$
for a non interacting tight-binding model, showing the doubly occupied one
particle states $\bar{k}$ (solid circles) in the one-particle energy
dispersion $e(\bar{k})$. Bottom: Energy as a function of flux of the
low-lying energy levels.}
\end{figure}

In previous cases, $l=1$ and the integration was restricted to one cycle of 
$\bar{H}$ ($0\leq \Phi \leq 2\pi $). However, while $K$ is kept fixed, 
$\bar{K}$ evolves from $K$ to $K+2\pi n/l$ in this cycle. 
Since when $l\neq 1$
these $\bar{K}$ are not equivalent, the initial and final states are
orthogonal and a Berry phase, even in its more general form \cite{sam},
cannot be defined for this circuit. In Fig. 1 I illustrate the evolution of
the ground state with $\Phi $ for $L=12$, $n/l=1/2$ for the case of a
non-interacting nearest-neighbor (NN) tight-binding Hamiltonian (Eq. (\ref
{fh}) for $U=V=0$). The minimum energy corresponds to $K=\Phi =0$. The
adiabatic continuation of this state in the representation of $H$ as $\Phi $
changes is always the same $|g_{0}(\Phi ,\Phi )\rangle =\prod_{\sigma }${${c}
_{-\pi /6\sigma }^{\dagger }{c}_{0\sigma }^{\dagger }{c}_{\pi /6\sigma
}^{\dagger }|0\rangle $, where ${c}_{k\sigma }^{\dagger }$ is the Fourier
transform of ${c}_{j\sigma }^{\dagger }$. In the representation of }$\bar{H}$
, each one-particle $\bar{k}$ is shifted by $\Phi /L$ with respect to the
corresponding $k$. It is easy to see that the energy of $|g_{0}(\Phi ,\Phi
)\rangle $ is $E_{0}(\Phi )=E(0)\cos (\Phi /L)$, where $E(0)$ is the energy
for $\Phi =0$ \cite{afq}. As $\Phi $ increases from 0 to $2\pi $, each 
$\bar{k}$ goes to the next allowed wave vector for PBC, 
and $|g_{0}\rangle $
evolves to an excited state with $\bar{K}=\pi $. For $\Phi $ near $2\pi $,
the ground state is $|g_{\pi }(\Phi ,\Phi )\rangle =\prod_{\sigma }{c}_{-\pi
/3\sigma }^{\dagger }${${c}_{-\pi /6\sigma }^{\dagger }{c}_{0\sigma
}^{\dagger }|0\rangle $ with }$K=\pi $. The crossing occurs for 
$\Phi=\pi $ and the ground state energy is of course periodic with period 
$2\pi $. 
This crossing does not affect the adiabatic theorem, since the states 
$|g_{0}\rangle $ and $|g_{\pi }\rangle $ have different  $K$. 
In contrast, $|g_{0}\rangle $ for {$\Phi =\pm (2\pi -0^{+})$ } 
represent orthogonal
eigenstates of $\bar{H}$ with the same $\bar{K}=0$ (see Fig. 1) and 
$\gamma_{c}$ is undefined (one of the factors in Eq. (1) vanishes). However, since
both states have the same quantum numbers, they should hybridize when
interactions are present, removing the degeneracy at $\Phi =\pm 2\pi $. In
general, for any $L$ and $n/l=1/2$, the above mentioned states differ in
four particles displaced from the neighborhood of the Fermi point $-\pi /4$
to near the other Fermi point $\pi /4$. Umklapp processes of this type are
generated in second and higher order perturbation theory in the interaction,
and were studied previously \cite{gia}. For other fractional occupancies the
situation is similar. The nature of the resulting thermodynamic phase as
well as the resulting value of $\gamma _{c}$ depends on the detailed form of
the interaction.

I have reexamined the relation between $\gamma _{c}$ and the macroscopic
polarization in the present case, following the derivation done previously
for $l=1$ in the many-body case \cite{or1,or2,niu}. Essential for the
extension is to work in a subspace with definite $K$, in such a way that for
all {$\Phi $ the ground state is separated from the excited states with the
same quantum numbers. I obtain that changes in polarization }$\Delta P$ are
related to the corresponding changes in $\gamma _{c}$ when some parameter of
the Hamiltonian varies by:

\begin{equation}
\Delta P=(e/2\pi l)\Delta \gamma _{c}\text{ (modulo }e/l).  \label{fp}
\end{equation}
The extra factor $l$ in the denominator is simply due to the $l$ times
larger interval of integration in the definition of $\gamma _{c}$. If $H(0,0)
$ has inversion symmetry, $H(\Phi _{\uparrow },\Phi _{\downarrow })$ is
transformed to $H(-\Phi _{\uparrow },-\Phi _{\downarrow })$ under inversion.
As a consequence $\gamma _{c}=-\gamma _{c}$ (modulo $2\pi $). This means
that in systems with inversion symmetry $\gamma _{c}=0$ or $\pi $ (modulo 
$2\pi $) as in previous cases \cite{zak1,mic,or2,ali1}.

The values of $\gamma _{c}$ and $\gamma _{s}$ are easy to predict for
systems and parameters in which the particles are localized or the relevant
kinetic terms are not affected by the flux. Examples are the CDW and SDW
phases with maximum order parameter at half filling, for which $\gamma
_{c}=\gamma _{s}=0$ and $\gamma _{c}=\gamma _{s}=\pi $ respectively \cite
{ali1}. Another examples which are of interest because of its competition
with S states, are states with phase segregation (PS) in which the particles
group together. Let us take a chain of $L=4N_{\sigma }$ sites in which the
first $N_{\sigma }$ ones are doubly occupied and the other are empty 
($n/l=1/2$). When Eq. (1) is applied to this state, all factors give 1, except
the last one which determines the phase: $\gamma _{c}=\sum
{}_{j=1}^{N_{\sigma }}8\pi j/L=\pi (N_{\sigma }+1)\equiv 0$ ($\pi $) if 
$N_{\sigma }$ is odd (even). The result is the same for states 
translated to
any other place in the chain, and to linear combinations of these states (in
particular with well defined $K$). This result suggests that choosing
appropriately $L$ one might be able to detect the boundary between S and PS
states. Another example in which $\gamma _{c}$ is easy to calculate is a CDW
for even $L$ in which every second state is singly occupied. Eq. (1) gives 
$\gamma _{c}=\sum {}_{i=1}^{L/2}8\pi i/L=\pi (L+2)\equiv 0$. The same result
is obtained if every fourth site is doubly occupied.

The fact that a Berry phase can jump sharply at the boundary between an S
phase and a CDW or PS state, can be useful in finite-size diagonalizations,
where correlation functions vary smoothly at the transition, and traditional
calculations, like those based on bosonization and conformal-field theory
results \cite{schulz}, which use the compressibility obtained numerically
from the energy necessary to add and remove two particles, may have large
finite-size effects. For example, in the phase diagram of the 1D generalized 
$t-J$ model including a three-site term, a superconducting bubble is
predicted inside the PS region, which is an artifact of finite-size effects
in the compressibility \cite{lem}. Instead, $\gamma _{c}$ does not depend on
the properties of the system for other densities and has a unique value on
the PS state. In other words, although a real phase transition for most
systems of interest can only take place in the thermodynamic limit, under
certain circumstances, both thermodynamic phases are characterized by
different topological quantum numbers, accessible in finite-size systems,
and with small size dependence. This was shown in Ref. \cite{ali1} for a
half filled system. Here I calculate $\gamma _{c}$ for the following 1D
model at quarter filling ($n/l=1/2$):

\begin{equation}
H(0,0)=-t\sum_{j\sigma }c_{j+1\sigma }^{\dagger }c_{j\sigma
}+U\sum_{j}n_{j\uparrow }n_{j\downarrow }+V\sum_{j\sigma \sigma ^{\prime
}}n_{j+1\sigma }n_{j\sigma ^{\prime }}.  \label{fh}
\end{equation}
I restrict most of the study to $U<0$. To distinguish the PS state from the
one with dominating singlet superconducting correlations at large distances
(S), it turns out that the odd $N_{\sigma }$ case mentioned above (leading
to $\gamma _{c}$(PS)$=0$) is the convenient choice. Thus, I took $L=12$, 
$N_{\uparrow }=N_{\downarrow }=3$, for the numerical calculations. As
expected from the above discussion, I obtain in the region $V>|U|\gg t$,
where charge-charge correlations dominate at large distances 
$\gamma _{c}$(CDW)$=0.$

\begin{figure}
\narrowtext
\epsfxsize=3.0truein
\vbox{\hskip 0.05truein \epsffile{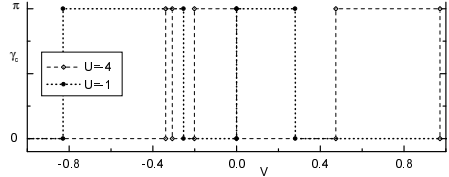}}
\medskip
\caption{Charge Berry phase as a function of $V$ for $U=-4$ (dashed line) and 
$U=-1$ (dotted line).}
\end{figure}

In the S region, $\gamma _{c}=0$ or $\pi $ depending on the parameters. Fig.
2 illustrates $\gamma _{c}$ as a function of $V$. There are several jumps in 
$\gamma _{c}$ which do not provide information on phase boundaries. In
particular, there is always a jump at $V=0$, and for large $|U|$ there is
another one near $V=V_{c}^{1}=-(t^{2}/|U|-4t^{4}/|U|^{3})$. To illustrate
this jump, consider the case $|U|\gg t,V$. In this limit, it is known that 
$H $ reduces to a model of $N_{\sigma }$ hard-core bosons, representing
doubly occupied sites, which can hop to NN positions with matrix elements 
$t^{\prime }e^{\pm i(\Phi _{\uparrow }+\Phi _{\downarrow })}$, with 
$t^{\prime }=2t^{2}/|U|$ ($t^{\prime }=-2V_{c}^{1}$ if fourth order
corrections are included), and NN repulsion $V^{\prime }=2(2V+t^{\prime })$.
In 1D, each hard-core boson can be mapped into a spinless fermion with an
appropriate change in the boundary conditions \cite{bat} and this model is
equivalent to an XXZ model $\sum_{<ij>\alpha }J_{\alpha }S_{i}^{\alpha
}S_{j}^{\alpha }$ with $N_{\sigma }$ spins up, and $J_{x}=J_{y}=2t^{\prime }$
, $J_{z}=V^{\prime }.$ Another way to reach the same result is to perform
first the unitary transformation ${c}_{j\uparrow }^{^{\prime }}=
{c}_{j\uparrow }$, ${c}_{j\downarrow }^{^{\prime }}=
(-1)^{j}{c}_{j\downarrow}^{\dagger }$ 
(which in $H$ has the effect of changing the sign of $U$ and 
$\Phi _{\downarrow }$ and transforming the NN repulsion in 
$4V\sum_{<ij>}S_{i}^{z}S_{j}^{z}$), and then eliminate terms linear in $t$
through a canonical transformation. For $V^{\prime }=0$ the model can be
solved trivially since the interaction vanishes in the spinless-fermion
model. Then, one might draw a picture like Fig. 1 to represent $|g_{0}(\Phi
,\Phi )\rangle $ and again there is a crossing of levels at $\Phi =\pm 2\pi $
of states with the same $K=0$, but different (singly) occupied (effective)
one-particle states $\bar{k}$, leading to an undefined $\gamma _{c}$.
However, in contrast to Fig. 1, states with $K=\pm \pi /2$ appear in the
low-energy manifold and the ground-state energy as a function of flux $\Phi $
displays periodicity in $\pi $ (with minima at $\Phi =\pi \times 
\mathop{\rm integer}$): 
the anomalous flux quantization characteristic of the S phase \cite
{ali1,afq,pen}. A small $V^{\prime }$ leads to a well defined $\gamma _{c}$
which depends on the sign of $V^{\prime }$. I also find other jumps of 
$\gamma _{c}$ for $V=0$ and at other points inside the S phase, which do not
provide information on phase boundaries. The physical meaning of the other
jumps inside the S phase is not obvious.

\begin{figure}
\narrowtext
\epsfxsize=3.0truein
\vbox{\hskip 0.05truein \epsffile{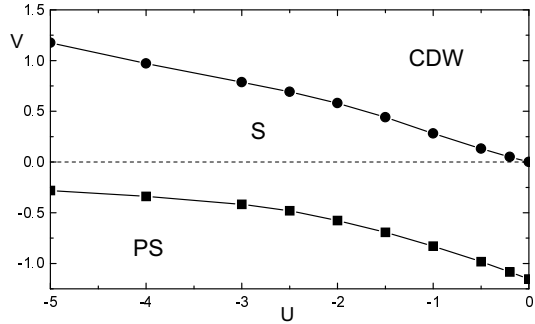}}
\medskip
\caption{Phase diagram of the extended Hubbard model Eq. (2) determined
through jumps in $\gamma _{c}$, showing the regions of dominating
superconducting (S) or charge-density wave (CDW) correlations at large
distance, and the region of phase segregation (PS).}
\end{figure}

Instead, I find that for $|U|\gg t$, the first and the last transition as a
function of $V$ agree very well with the limits of the S phase, obtained in
the thermodynamic limit through the exact Bethe-ansatz solution of
equivalent XXZ models: from the solution of the above mentioned XXZ model 
\cite{fow}, we know that PS occurs for $J_{z}<-J_{x}$ (ferromagnetic
Ising-like case), what in terms of the parameters of Eq. (\ref{fh}) means 
$V<V_{PS}=-2t^{2}/|U|$. On the other hand, when $V\gg t^{2}/|U|$, extending
previous ideas \cite{bat,pen}, the entity composed of a doubly occupied site
and an empty site right to it can be mapped into a spinless fermion. After
an appropriate canonical transformation \cite{pen}, I find that the
resulting spinless fermion model has on-site energy $U-4t^{2}/(V-U)$, NN
repulsion $V^{\prime }=4t^{2}[1/(V-U)-1/(3V-U)]$, hopping $t^{\prime }e^{\pm
i(\Phi _{\uparrow }+\Phi _{\downarrow })}$, with $t^{\prime }=2t^{2}/(V-U)$,
number of particles $N^{\prime }=N_{\sigma }=N/2$ and number of sites 
$L^{\prime }=L-N/2$. I have calculated the correlation exponent $K_{\rho }$
of this model in the thermodynamic limit from its compressibility and charge
velocity $v_{c}$ mapped appropriately to the original model \cite{bat}. The
energy and $v_{c}$ were obtained solving the corresponding integral
equations of the equivalent XXZ model \cite{fow}. I obtain that the boundary 
$K_{\rho }=1$ between dominating S or CDW correlations at large distances
corresponds to $J_{z}\cong 0.25J_{x}$, leading to $V_{CDW}\cong 0.20|U|$.
For $5t<|U|<10t$, I find that for $L=12$, the first jump in $\gamma _{c}$
with increasing $V$ is above $V_{PS}$ by $\sim 25\%$ and the last jump is
above $V_{CDW}$ by $\sim 15\%$. These numbers are reduced by a factor $\sim
2/3$ if fourth order corrections to $t^{\prime }$ and $V^{\prime }$ are
included. I believe that the remaining discrepancy is due to finite size
effects. For smaller values of $|U|$, the first and last jumps in 
$\gamma_{c}$, shown in Fig. 3 agree with previous numerical calculations of 
$K_{\rho }$ and compressibility \cite{pen,sano}. In correspondence with these
calculations, I find an island for high $V$ and small $|U|$ ($V\sim 8t$, 
$U\sim -2t$, not shown) in which $\gamma _{c}$ jumps to $\pi $ and $K_{\rho
}>1$. However, using again an appropriate mapping to two XXZ models, I
obtain that this is a finite-size effect and that in the thermodynamic limit
the system phase separates into two phases with dominating CDW correlations,
in agreement with previous suggestions \cite{pen}. One phase is the one just
described for $V>V_{CDW}$, and in the other one fermion objects composed of
one particle and an empty site next to it move with hopping $t$ and NN
attraction $V^{\prime }=-2t^{2}/V$.

Another limit in which the charge dynamics is equivalent to that of
interacting spinless fermions is that of $U=+ \infty$. Following similar
methods as those already used for $V=0$ \cite{cas}, the model can be mapped
into the spin polarized case $N_{\uparrow }=L/2$, $N_{\downarrow }=0$, with
a shift $\Delta \Phi$ in the BC. Calculating $K_{\rho }$ in the
thermodynamic limit, I find that superconducting correlations dominate for 
$-2t<V<V_{CDW}$, where up to three digits accuracy $V_{CDW}=-\sqrt{2}t$. For 
$8\leq L\leq 16$, $\gamma _{c}$ jumps exactly at this value . Instead, 
$\gamma _{c}$ does not detect the transition to the PS state at $V=-2t$ in
this case.

In the rest of this Letter, I discuss the spin Berry phase $\gamma _{s}$. As
noted earlier \cite{ali1,ali2}, $\gamma _{c}$ is transformed into $\gamma
_{s}+\pi $ (of a different Hamiltonian in general) and vice versa by the
transformation ${c}_{j\uparrow }^{^{\prime }}={c}_{j\uparrow }$, 
${c}_{j\downarrow }^{^{\prime }}=(-1)^{j}{c}_{j\downarrow }^{\dagger }$. 
Thus,
the definition of $\gamma _{s}$ can be generalized to $N_{\uparrow }\neq
N_{\downarrow }$ in the same way as $\gamma _{c}$, keeping $\Phi
_{\downarrow }=-\Phi _{\uparrow }$ in the $l$ cycles, and $l$ should be such
that $(N_{\uparrow }-N_{\downarrow })l/L$ is an integer. It is also clear
from this transformation and what I obtained for $\gamma _{c}$ (Eq.(\ref{fp}
)) that $\Delta (P_{\uparrow }-P_{\downarrow })=(e/2\pi l)\Delta \gamma _{s}$
(modulo $e/l),$where $P_{\sigma }$ is the macroscopic polarization of the
particles with spin $\sigma $. If $N_{\uparrow }=N_{\downarrow }$, as in
most cases of interest, the previous definition of $\gamma _{s}$ with $l=1$ 
\cite{ali1}, is valid for any filling, and for 1D systems with spin SU(2)
symmetry, $\gamma _{s}$ jumps in $\pi $ when a spin gap $\Delta _{s}$ opens.
To show this, I use recent results from continuum limit theory and
bosonization \cite{nak}, which show that the opening of a spin gap is
determined by the crossing of the lowest energy states of $\bar{H}$ with
total spin $S=0$ and $S=1$ within the $\bar{K}=0$ sector, for periodic
(antiperiodic) BC if $N_{\sigma }$ is even (odd). These BC correspond to a
particular point $\Phi =\Phi _{c}$ with $\Phi _{c}=0$ ($\Phi _{c}=\pi $) in
the trajectory with $\bar{K}=K=0$ of $H(\Phi ,-\Phi )$ used to define 
$\gamma _{s}$. Usually this point corresponds to the {\em maximum} of the
ground-state energy $E_{0}(\Phi ,-\Phi )$ as a function of $\Phi $, while
its minimum lies at $\Phi =\pi $ ($\Phi =0$) for $N_{\sigma }$ even (odd),
as expected from the non-interacting limit. 

From its definition, it is obvious that to calculate $\gamma _{s}$ any
(single valued) $\Phi $-dependent gauge can be chosen. I take the
representation in terms of the operators $\tilde{c}_{j\uparrow }^{\dagger
}=e^{ij\Phi _{c}/L}{c}_{j\uparrow }^{\dagger }$, $\tilde{c}_{j\downarrow
}^{\dagger }=e^{-ij\Phi _{c}/L}{c}_{j\downarrow }^{\dagger }$, in such a way
that in the resulting Hamiltonian $\tilde{H}(\Phi ,-\Phi )$, all hoppings
are real for $\Phi =\Phi _{c}$ (as in $\bar{H}(\Phi _{c},-\Phi _{c})$ ).
Sufficiently near to $\Phi _{c}$, one can approximate $\tilde{H}(\Phi ,-\Phi
)\simeq \tilde{H}(\Phi _{c},-\Phi _{c})+\frac{\partial \tilde{H}}{\partial
\Phi }(\Phi -\Phi _{c})$, where the derivative is evaluated at $\Phi _{c}$.
While at $\Phi =\Phi _{c}$, the SU(2) symmetry is conserved, for $\Phi \neq
\Phi _{c}$ only spin rotations around the quantization axis $z$ are
conserved by $\tilde{H}$. Specifically, $\partial \tilde{H}/\partial \Phi $
is equal to the paramagnetic part of the difference between up and down
currents, and changes sign under spin rotations in $\pi $ around the $x$ or 
$y$ axis. I denote these rotations by $R$ and call $|s\rangle $, $E_{s}$ (
$|t\rangle $, $E_{t}$) the lowest singlet (triplet with spin projection zero)
eigenstate of $\tilde{H}(\Phi _{c},-\Phi _{c})$ and its energy. It is known
that $R|s\rangle =|s\rangle $, and $R|t\rangle =-|t\rangle $. Since 
$\partial \tilde{H}/\partial \Phi $ is odd under $R$, this implies $\langle
s|\partial \tilde{H}/\partial \Phi |s\rangle =\langle t|\partial \tilde{H}
/\partial \Phi |t\rangle =0$, However, in general $\langle s|\partial 
\tilde{H}/\partial \Phi |t\rangle =A\neq 0$, where $A$ can be made real by a
suitable change of the phase of one of the states. Near the crossing point
between lowest singlet and triplet energies, the ground state $|g_{0}(\Phi
,-\Phi )\rangle $ results from the diagonalization of the low-energy part 
$\tilde{H}_{LE}$ of $\tilde{H}$ containing the states $|s\rangle $ and 
$|t\rangle $, which according to the above symmetry arguments, has the form:

\begin{equation}
\tilde{H}_{LE}(\Phi ,-\Phi )\simeq \left( 
\begin{array}{ll}
E_{S} & A(\Phi -\Phi _{c}) \\ 
A(\Phi -\Phi _{c}) & E_{T}
\end{array}
\right) .  \label{fm}
\end{equation}
It is easy to see that this leads to a jump in $\pi $ in $\gamma _{s}$ when 
$E_{T}$ crosses $E_{S}$. For example, one can exploit again the gauge
invariance to render $\langle s|g_{0}(\Phi ,-\Phi )\rangle $ real and
positive for $\Phi $ near $\Phi _{c}$, leading to $|g_{0}(\Phi
_{c}+0^{+},-\Phi _{c}-0^{+})\rangle =|g_{0}(\Phi _{c}-0^{+},-\Phi
_{c}+0^{+})\rangle $ for $E_{S}<E_{T}$ but $|g_{0}(\Phi _{c}+0^{+},-\Phi
_{c}-0^{+})\rangle =-|g_{0}(\Phi _{c}-0^{+},-\Phi _{c}+0^{+})\rangle $ for 
$E_{S}>E_{T}$ and no other contribution to $\gamma _{s}$ in the vicinity of 
$\Phi _{c}$. I obtain numerically in the extended Hubbard model with or
without correlated hopping \cite{ali1} that $\gamma _{s}=0$ in the spin
gapped phase and therefore $\gamma _{s}=\pi $ if $\Delta _{s}=0$. According
to continuum limit theory and renormalization group, in 1D, when $K_{\rho
}>1 $, the opening of a spin gap signals the transition from dominating
triplet to singlet superconducting correlations at large distance.

In summary, I have shown the ability of two Berry phases, and topological
quantum numbers derived from them, to detect phase transitions, particularly
in strongly correlated systems with non-integer number of particles per
site, for which no previous applications of Berry phases were considered. I
hope that this study will stimulate further research and applications of the
Berry phases in phase transitions, particularly in dimensions higher than
one, where alternative methods which use results of conformal field theory,
or exact solutions \cite{afq,gia,lem,bat,pen,fow,sano,nak} are not available.

I am indebted to Gerardo Ortiz for providing me with details of his
calculations in Ref. \cite{or1} and useful discussions. The numerical
diagonalization was done using subroutines developed by Eduardo Gagliano,
who unfortunately died on February 12, 1998. I am partially supported by
CONICET, Argentina.

\end{document}